\documentclass[journal,comsoc]{IEEEtran}

\usepackage[T1]{fontenc}
\usepackage{graphicx}
\usepackage{cite}
\usepackage{subcaption}
\usepackage{amsmath,amssymb}
\usepackage{caption}
\usepackage{booktabs}
\usepackage{array}
\newcolumntype{M}[1]{>{\centering\arraybackslash}m{#1}}
\newcolumntype{L}[1]{>{\raggedright\arraybackslash}m{#1}}
\usepackage{multirow}
\usepackage[table]{xcolor}

\title{LLM-Enhanced Space-Air-Ground-Sea Integrated Networks}

\author{Halvin~Yang,~Sangarapillai~Lambotharan,~Mahsa~Derakhshani,~Lajos~Hanzo\\
	\thanks{Halvin Yang and Mahsa Derakhshani are with the Wolfson School of Mechanical, Electrical and Manufacturing Engineering, Loughborough University, UK.}
	\thanks{Sangarapillai Lambotharan is with the Institute for Digital Technologies, Loughborough University, UK.}
	\thanks{Lajos Hanzo is with the University of Southampton, UK.}
	\thanks{Corresponding author: Lajos Hanzo (email: lh@ecs.soton.ac.uk).}}

\begin{document}
	
	\maketitle
	
	\begin{abstract}
		The space–air–ground–sea integrated networking (SAGSIN) concept promises seamless global multimedia connectivity, yet two obstacles still limit its practical deployment. Firstly, high-velocity satellites, aerial relays and sea-surface platforms suffer from obsolete channel state information (CSI), undermining feedback-based adaptation. Secondly, data-rate disparity across the protocol-stack is extreme: terabit optical links in space coexist with kilobit acoustic under-water links. This article shows that a single large language model (LLM) backbone—trained jointly on radio, optical and acoustic traces—can provide a unified, data-driven adaptation layer that addresses both rapid CSI ageing and severe bandwidth disparity across the SAGSIN protocol-stack. Explicitly, an LLM-based long-range channel predictor forecasts the strongest delay–Doppler components several coherence intervals ahead, facilitating near-capacity reception, despite violent channel fluctuations. Furthermore, our LLM-based semantic encoder turns raw sensor payloads into task-oriented tokens. This substantially reduces the SNR required for high-fidelity image delivery in a coastal underwater link, circumventing the data rate limitation by semantic communications. Inclusion of these tools creates a medium-agnostic adaptation layer that spans radio, optical and acoustic channels. We conclude with promising open research directions in on-device model compression, multimodal fidelity control, cross-layer resource orchestration and trustworthy operation, charting a path from laboratory prototypes to field deployment.
		
	\end{abstract}
	
	\begin{IEEEkeywords}
		space–air–ground–sea integrated network, large language model, channel prediction, semantic communication, fluid antenna, 6G
	\end{IEEEkeywords}
	
	\section{Introduction}

	Early 6G research converges on a challenging strategic goal: ubiquitous, high-capacity connectivity that persists, regardless whether the endpoint is in a dense city, a remote desert, the open ocean or in low-Earth orbit. Achieving this goal requires a network footprint far wider—and far more flexible—than what today’s fibre-anchored terrestrial base-station grid can support.
	
	Progress with 5G non-terrestrial networks has narrowed the coverage gap \cite{Saad2024}, yet global reach still hinges on terrestrial infrastructure. Fibre backhaul is commercially untenable in deserts, polar regions and deep oceans, leaving shipping lanes, offshore platforms and research stations without dependable broadband connectivity. Even where coverage exists, the links on fast-moving platforms such as LEO satellites or airplanes may falter when the channel-state information (CSI) is outdated.
	
\begin{figure*}[t]
		\centering
		\includegraphics[width=0.7\textwidth]{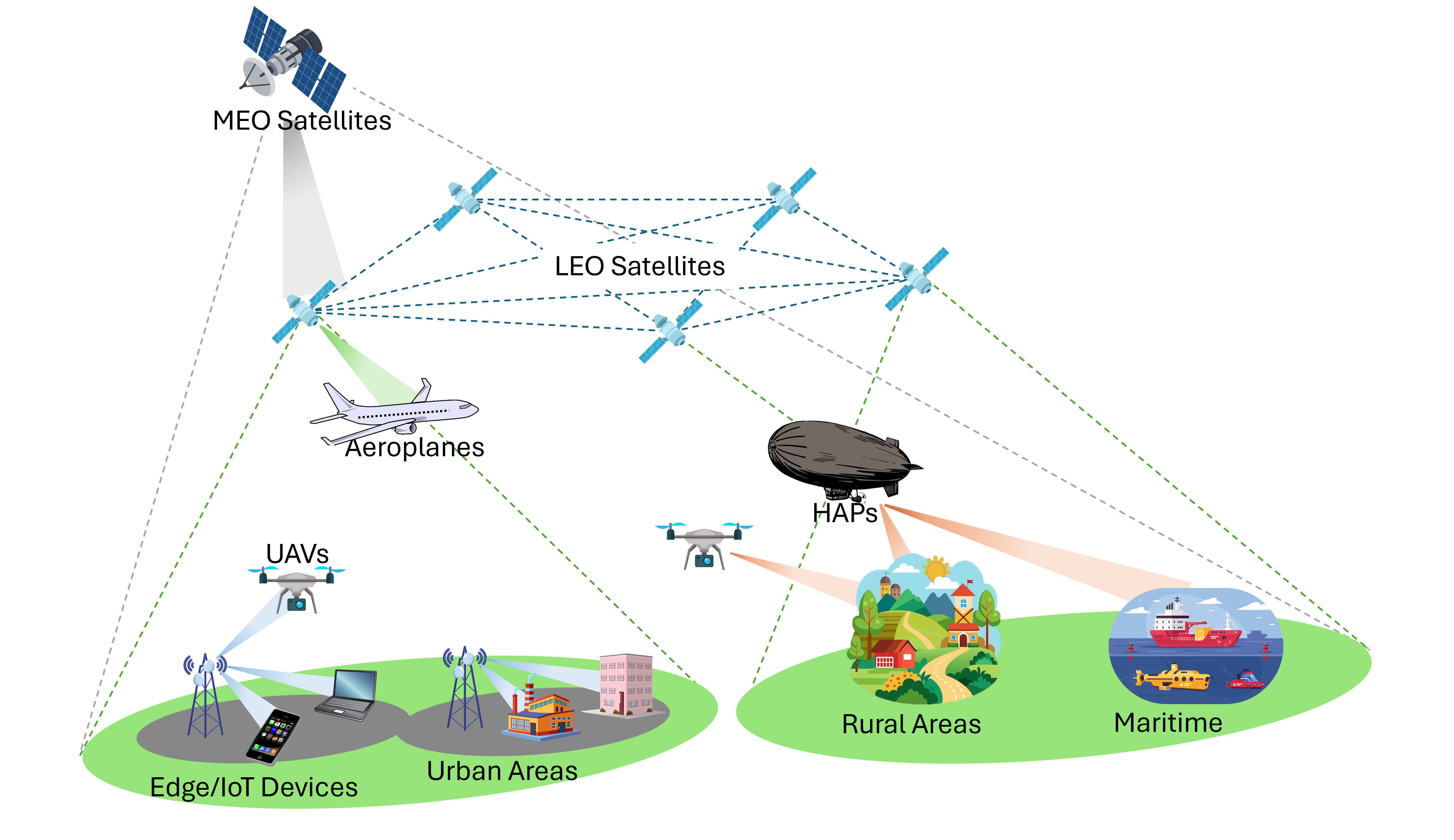}
		\caption{An example of how Space-Air-Ground-Sea networks can be connected in a SAGSIN.}
		\label{fig:sagsin}
	\end{figure*}

	Figure~\ref{fig:sagsin} depicts a space–air–ground–sea integrated network (SAGSIN) that unifies four operational layers \cite{Guo2022}. Low earth orbit (LEO) satellite constellations provide globe-spanning backhaul; high-altitude platforms (HAPs, 18--25 km) and unmanned aerial vehicles (UAVs) fill access gaps and extend free-space-optical trunks; terrestrial cells guarantee dense horizontal coverage; and maritime surface gateways, buoys and underwater modems provide services into offshore and subsea domains. Adding the space and sea strata extends seamless connectivity to areas that fibre-centric infrastructure cannot serve economically.
	
	That architectural promise comes with formidable obstacles. High mobility in the space layer yields Doppler excursions above 40 kHz at Ku-band and wide delay–Doppler spreads. Sea-surface optical and underwater acoustic links face acute bandwidth (kilobits per second) and power limits. Across every layer, heterogeneous media—Ka/Ku radio, millimetre- and terahertz-wave, free-space optics and acoustics— undermine any hope for a single waveform, coding scheme or resource-allocation routine.
	
	Large language models (LLMs) can address these issues from a data-driven angle. Compared to earlier data-driven approaches like Kalman filtering, CNN/RNN predictors or graph neural networks, transformer-based LLMs offer three capabilities that are uniquely valuable for SAGSIN.
	Firstly, self-attention projects heterogeneous inputs (radio IQ samples, optical CSI matrices, acoustic spectrograms) into a shared token space, so that a single backbone can span all four SAGSIN layers \cite{Zhou2024, 10558819}.
	Secondly, attention windows extend over hundreds of tokens, allowing the model to capture long-horizon temporal correlations and forecast several coherence intervals ahead—something recurrent nets struggle with because of vanishing gradients \cite{Yang2025}.
	Thirdly, LLMs can be retargeted on the fly through in-context prompts or lightweight low-rank adaptation (LoRA) fine-tuning, avoiding expensive full retraining when spectrum, mobility or traffic priorities change \cite{Hu2021}.
	These properties make LLMs a natural unifying “brain” for heterogeneous, rapidly fluctuating space–air–ground–sea links.

	For example, transformers have forecast near-optimal millimetre-wave and terahertz beams by attending to past beam indices \cite{Zhou2024}, inferred power-control and spectrum-allocation vectors from textual prompts, and driven knowledge-free network optimisations via GPT-style agents. Prompt-chain search has even outperformed heuristics for large-scale access-point placement. These results show that transformer backbones can learn and act on wireless-specific patterns directly from raw data, without relying on explicit channel models.
	
	Yet two research gaps remain. Firstly, no unified LLM-based channel-prediction framework operates seamlessly across all SAGSIN layers, where extreme Doppler and mixed spectra prevail. Secondly, the potential of LLM-assisted semantic communication to ease the grave data-rate disparity between space trunks and sea links is still under-explored. We argue that combining LLM-powered prediction with task-aware semantic coding provides the missing cross-layer cohesion.
	
	The remainder of this article develops these solutions. Section~2 details the SAGSIN architecture and quantifies its principal technical challenges. Section~3 introduces a fluid-antenna-aware LLM predictor for proactive CSI estimation throughout the four-layer stack. Section~4 presents an LLM-driven semantic-coding pipeline tailored to SAGSIN’s most bandwidth-constrained links. Section~5 discusses open issues—cross-layer resource optimisation, multimodal semantics and model compression—while Section~6 concludes with a roadmap toward practical deployment.

	
	\section{High-Level Overview of Space–Air–Ground–Sea Communications}
\begin{figure*}[t]
		\centering
		\includegraphics[width=1\textwidth]{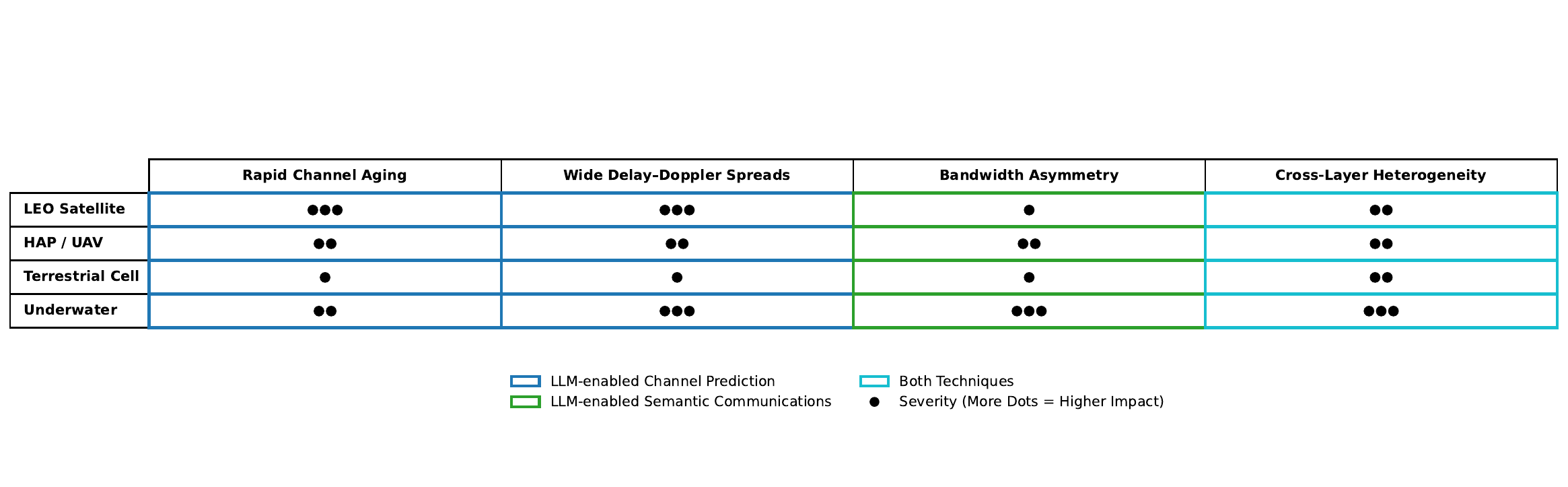}
		\caption{Overview of SAGSIN and its limitations.}
		\label{fig:limitations}
	\end{figure*}

	This section presents recent developments in SAGSIN and discusses several limitations that have to be addressed. These are summarised in Figure \ref{fig:limitations}.
	
	\subsection{Layered Topology and Representative Prior Work}
	
	A space–air–ground–sea integrated network (SAGSIN) assembles four strata whose propagation and hardware budgets differ so gravely that a design proven in one tier rarely applies to the next. A concise, layer-by-layer snapshot sets the stage for the technical challenges that follow.
	
	\begin{itemize}
		\item \textbf{Space layer (LEO/MEO–GEO).} Constellations of low-Earth-orbit satellites provide globe-spanning connectivity—serving both as high-capacity back-haul for remote gateways and, increasingly, as direct-to-device access links—with one-way delays of roughly 15–45 ms. Their 7 km/s orbital velocity, however, pushes Ku-band Doppler shifts beyond 40 kHz. Recent progress shows that LEO constellations have outgrown their early “back-haul only” role. Demonstrations by commercial systems now deliver broadband and LTE-NB-IoT links directly to handsets, while multi-gigabit optical inter-satellite links enable space-borne mesh routing that cuts end-to-end latency. These capabilities are being codified in 3GPP Releases 17–18, which specify non-terrestrial network (NTN) random-access, timing-advance and hybrid automatic repeat request (HARQ) procedures for both LEO and MEO data regimes, albeit not for real-time speed. Collectively, they elevate the space tier from a passive relay to an adaptive layer that participates in real-time optimisation across the entire SAGSIN stack \cite{Saad2024}. 
		
		\item \textbf{Air layer (HAPs and UAVs).} High-altitude platforms at 18–25~km and fixed- or rotary-wing UAVs act as agile gap-fillers, provide free-space-optical trunks and enable near-space relays. A recent survey framed near-space communications as the last piece in the SAGSIN stack \cite{Liu2024}, underscoring open questions in platform placement and cross-tier hand-over.
		
		\item \textbf{Ground layer.} Terrestrial macro, micro and pico cells—backed by edge data centres—anchor traffic aggregation and local processing. Recent research therefore concentrates on the interfaces between terrestrial cells and non-terrestrial assets, supporting seamless inter-tier hand-over procedures, flexible spectrum-sharing frameworks, and traffic-steering algorithms that divert bursts to UAV or LEO relays when ground cells approach congestion. 
		
		\item \textbf{Sea layer.} Surface gateways on ships or buoys relay RF/FSO traffic, while optical or acoustic modems extend links to autonomous underwater vehicles and subsea sensor swarms. Hybrid RF–optical links and adaptive acoustic modulation schemes define the current state of practice \cite{Wang2021}.
	\end{itemize}
	
	Beyond single-tier analyses, several multi-tier studies have emerged. A federated-learning framework examined distributed model training across all four layers, revealing security and privacy risks in heterogeneous media. Another overview introduced semantics-empowered SAGSINs \cite{Meng2025}, arguing that meaning-aware source coding becomes vital when the bandwidth spans five orders of magnitude; from 100~Gb/s optical trunks to kilobit-level acoustic channels. Collectively, these studies confirm the architectural feasibility of SAGSIN, while exposing persistent bottlenecks.
	
	\subsection{Outstanding Limitations}
	
	\subsubsection*{\textbf{Rapid channel ageing}}
	Orbital motion, UAV manoeuvres and sea-surface dynamics alter channel coefficients on millisecond scales, rendering feedback-driven adaptation obsolete before new settings take effect. A robust, media-agnostic method for forward channel prediction is still lacking.
	
	\noindent For concreteness, we consider Orthogonal Time Frequency Space (OTFS) modulation—a delay–Doppler waveform expressly designed for high-mobility links. As an example, a LEO satellite travelling at 7 km/s can induce Doppler shifts exceeding 40 kHz within a single OTFS modulation frame, rendering any feedback-based CSI adaptation obsolete before it can be applied. Similarly, underwater acoustic channels may decorrelate within tens of milliseconds due to wave motion and temperature gradients. In such environments, without forward-looking prediction, beam alignment and port selection rapidly become outdated.
	
	\subsubsection*{\textbf{Wide delay–Doppler spreads}}
	High mobility in LEO links and rich multipath in shallow-water acoustics broaden delay–Doppler range, inflate guard intervals and weaken coherent detection.
	
	\noindent In practice, Doppler spreads in LEO Ka-band links can reach 7–10 kHz, while shallow-water acoustics may exhibit delay spreads of several milliseconds. These broadened regions require longer cyclic prefixes or guard intervals, which in turn reduce spectral efficiency and challenge standard coherent detectors that assume narrowband approximations.
	
	\subsubsection*{\textbf{Severe data-rate asymmetry}}
	Inter-satellite optical trunks exceed 100~Gb/s, whereas acoustic sea links may struggle to reach a few kilobits per second under tight power budgets. Such disparity calls for task-oriented compression rather than bit-exact coding.
	
	\noindent A clear disparity arises when, for instance, a surface platform has to forward high-resolution sonar data to a satellite via a 2 kbps underwater link. Even lossy compression becomes impractical. Bit-exact transmission of such content across mismatched links leads to intolerable delays and packet losses, reinforcing the need for task-specific semantic representation.
	
	\subsubsection*{\textbf{Cross-layer media heterogeneity}}
	No single waveform or code family excels simultaneously over Ka/Ku radio, mmWave/THz, FSO and acoustic channels, so protocol translation at gateways adds latency and complexity.
	
	\noindent A mission-critical packet may have to traverse a sequence of media—starting with a Ku-band RF hop from a satellite, moving through a millimeter-wave HAP–UAV link, and ending with an underwater acoustic segment. Each of these imposes unique channel impairments and bandwidth constraints, yet no single waveform or protocol can perform optimally across all of them. As a result, inter-layer gateways often introduce format conversion delays, additional latency, and loss of synchronization. These transitions demand intelligent, context-aware translation mechanisms that go beyond static protocol stacks.
	
	\medskip
	Large language models offer data-driven remedies for the first and third limitations. Transformer-based predictors have already outperformed Kalman filters and recurrent networks in satellite and aerial trials, heralding their promise for unified channel prediction across the SAGSIN stack. This is because their self-attention mechanism can capture long-range temporal dependencies, fuse heterogeneous features from radio, optical and acoustic traces into a common embedding, and adapt on-the-fly via lightweight fine-tuning. These are advantages that conventional Kalman filters or recurrent nets lack in doubly selective, multi-modal SAGSIN links. Likewise, LLM-enabled semantic communication has delivered order-of-magnitude SNR savings for high-resolution imagery, indicating a practical path to easing bandwidth constraints—especially on acoustic underwater links. Section~3 therefore introduces a fluid-antenna-aware LLM predictor for proactive CSI estimation, while Section~4 presents an LLM-driven semantic-coding pipeline tailored for SAGSIN’s most bandwidth-restricted links.

	\section{LLM-Aided Channel Prediction for SAGSIN}
	
	\subsection{Channel modelling and two-stage compression}
	A Ka-band LEO satellite, for example, sweeps over a coastal terminal at roughly 7\,km/s, imposing a 40 kHz Doppler shift within a single OTFS frame, while an underwater acoustic-RF gateway endures rapidly drifting surface reflections and microbubble scattering. Such mobility, dispersion and intermittency create a four-dimensional channel that can change appreciably during the time it would take a pilot-only scheme to acquire full CSI. Channel prediction therefore becomes mandatory: forecasting the next few frames allows beamforming, modulation and scheduling decisions to remain valid throughout their execution window. We build on the FAS-LLM framework proposed in \cite{Yang2025_FASLLM}, which couples a large language model with fluid antenna systems whose reconfigurable antenna ports can be switched to the specific location in space having the highest signal quality. Port agility boosts diversity and outage resilience, yet it also introduces an extra spatial dimension that the predictor must learn—identifying not only how delay and Doppler will evolve, but which port will dominate in the forthcoming frame.

	To address this, a lightweight two-stage compression pipeline distils the matrix into a compact, information-rich token stream. The first stage, namely reference-port selection, exploits the fact that fluid antenna ports differ mostly in their deterministic phase ramps. During a brief calibration, the receiver identifies the specific port having the strongest average power; all subsequent frames are represented only by that port, while the others are later regenerated analytically. The second stage performs separable principal-component analysis along the spatial and combined delay–Doppler axes. A small set of orthogonal patterns on each axis captures the majority of the received energy. Projecting the reference-port matrix onto these patterns reduces each frame to a few dozen real numbers that are normalised, quantised and tokenised for the LLM. Reconstruction after prediction is deterministic: the coefficients are expanded through the two bases; the port-dependent phase offsets are reapplied, and the full four-dimensional channel of the next frame is recovered. This channel-model, matrix and compression pipeline reconciles rich physical-layer dynamics with the strict token budget of modern LLMs.
	
	\subsection{Experimental scenario and implementation}
	The evaluation considers a demanding yet representative scenario in which an eight-by-eight Ka-band array on a LEO satellite at 600km serves a sixteen-port linear fluid antenna mounted on a coastal IoT buoy. The waveform is off-grid OTFS used in prior work, employing a sub-carrier spacing of 1 kHz on a 64 $\times$ 64 delay–Doppler lattice. The maximum Doppler spread reaches 7 kHz, and six dominant Ricean paths carry most of the energy. Fifty past frames form the look-back window, and the predictor forecasts twenty future frames, remaining within the measured coherence interval.
	
	A $1 \times 10^9$ LLaMA-3 backbone is fine-tuned using low-rank adaptation. Only two narrow linear heads and the LoRA modules are updated, keeping the trainable parameter count below $2 \times 10^6$. Training completes in $< 10$ GPU-hours. A convolutional transformer and GPT-2 serve as baselines, all trained with identical compressed tokens, data and optimisation settings.
	
	\subsection{Performance evaluation and discussion}
	
	\begin{figure}[t]
		\centering
		\includegraphics[width=1\linewidth]{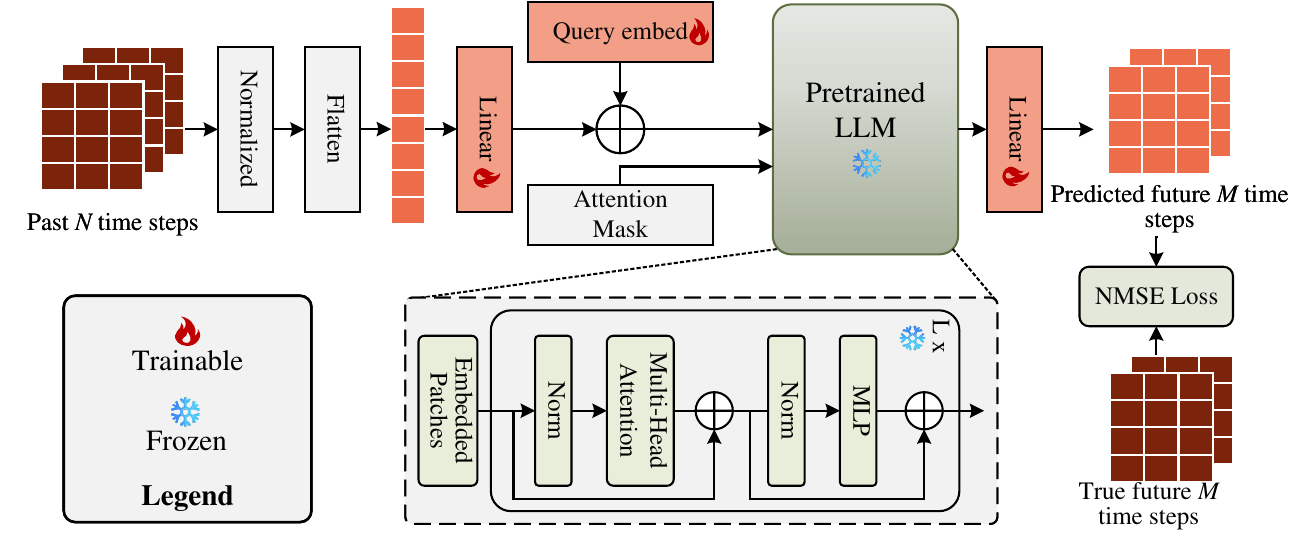}
		\caption{LoRA-adapted LLM architecture with compressed channel inputs for multistep prediction across SAGSIN links.}
		\label{fig:LLM_arch}
	\end{figure}
	
	Figure~\ref{fig:LLM_arch} illustrates the five-stage pipeline that forecasts future channels from past snapshots. Raw delay–Doppler tensors are first compressed via reference-port selection and separable PCA, preserving 90\% of the input energy, while shrinking the problem dimensionality bto $<1\%$. The resultant 64 coefficients are $\Delta$-encoded, quantised to 8-bit integers, and tokenised after per-frame normalisation using byte-pair encoding aligned to the LLaMA-3 vocabulary.\footnote{Compression uses a rank-$(4,16)$ basis across spatial and delay–Doppler dimensions, retaining 90\% signal energy as measured by the Frobenius norm across 1,000 frames. Tokenisation converts the quantised coefficients into 96 tokens plus a query token, using the LLaMA-3 vocabulary.} These 97 tokens are passed through frozen LLaMA-3 transformer blocks equipped with rank-8 LoRA adapters \cite{Devalal2018}, and the model is trained to minimise the normalised mean-square error contaminating the compressed channel vector.\footnote{Training loss is NMSE computed on the 64-dimensional compressed vector, averaged over 1,000 Monte-Carlo runs.} A deterministic PCA expansion followed by port phase ramping reconstructs the full-resolution channel matrix for downstream use.

	\begin{figure}[t]
		\centering
		\includegraphics[width=1\linewidth]{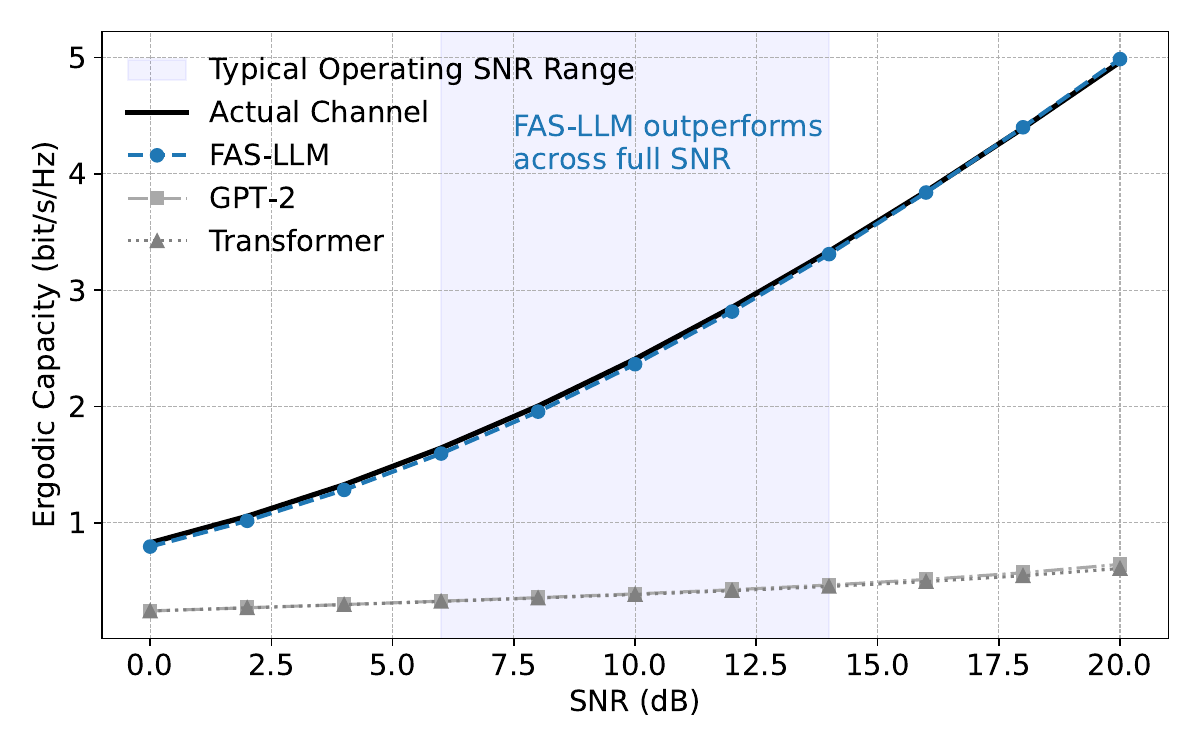}
		\caption{Ergodic capacity versus signal-to-noise ratio for predicted and actual channels using a twenty-frame prediction horizon. The proposed LLM predictor approaches the perfect CSI bound, while gated-recurrent units, long short-term memory networks, a convolutional transformer and GPT-2 incur significant capacity loss.}
		\label{fig:capacity}
	\end{figure}
	
	Figure~\ref{fig:capacity} illustrates the system-level benefits of accurate long-range channel prediction, including a lightly shaded band (5–14 dB) that marks the typical operating SNR range for SAGSIN links. The FAS-LLM model consistently follows the actual channel capacity curve, maintaining a near-optimal match across the SNRs of 0-20 dB, never diverging by more than 0.03 bit/s/Hz. Inside the highlighted operating band, this translates into less than a 1\% capacity penalty, ensuring that proactive beam-switching and port-selection decisions are made with near-optimal accuracy. By contrast, GPT-2 and a convolutional transformer deliver barely one quarter of the attainable throughput even at the highest SNR tested, and fall well below the 1 bit/s/Hz threshold throughout most of the practical range. 
	This confirms that once the channel snapshot is distilled by the two-stage compression front-end, a lightweight, LoRA-aided LLM can forecast future channels with sufficient fidelity to approach perfect-CSI based performance across the SNR envelope of interest. A LLM can accurately capture channel dynamics and enable reliable communication without excessive feedback overhead or adaptation delay.

	\section{LLM-Assisted Semantic Communications for SAGSIN}\label{section:SC}
	
	\subsection{Method}
	Semantic communication (SC) \cite{10558819} redefines the paradigm of communication by shifting the objective from transmitting raw data to conveying task-specific semantics. In contrast to traditional systems that prioritize bit-level fidelity, SC emphasizes the extraction, compression, and transmission of high-level semantic features essential for task execution, leveraging advanced AI models. This approach effectively filters out redundant information, substantially reducing data volume and operating efficiently under bandwidth constraints\footnote{Source data in our experiments comes from the full SAGSIN stack: LEO satellite imagery, high-altitude snapshots, coastal camera feeds, and subsea photographs.}.
	
	However, dynamically fluctuating SAGSIN channels tend to force traditional SC systems based on lightweight neural networks into frequent retraining or update models to maintain semantic alignment and task accuracy. These frequent updates are further exacerbated by the well-known problem of memory-erosion in conventional neural networks, where previously learned knowledge is rapidly overwritten when adapting to new domains. This severely limits the adaptability and scalability of traditional SC solutions.
	
\begin{figure}[htbp]
		\centering
		\includegraphics[width=8.5cm]{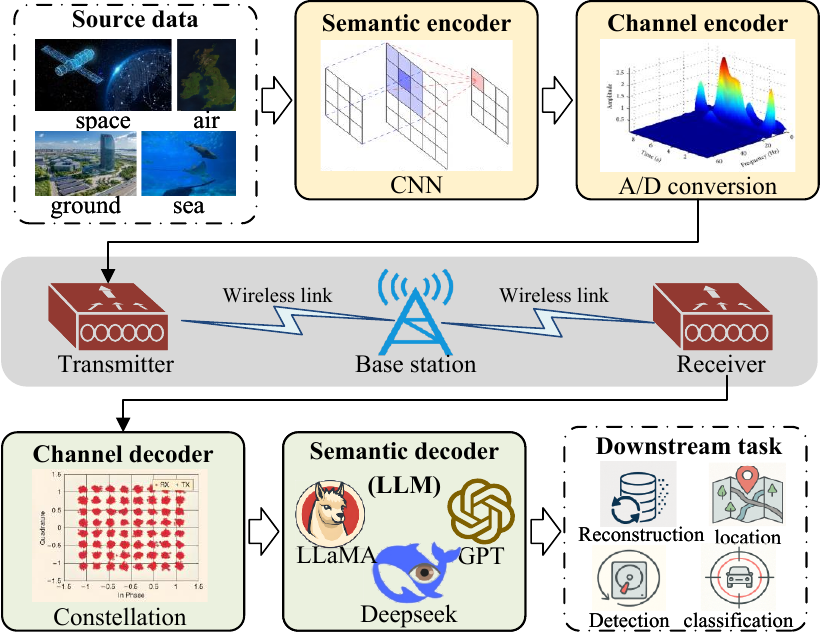}
		\caption{The illustration of the LLM-assisted SC model.}
		\label{fig:semcom}
	\end{figure}

	To overcome these limitations, we propose an LLM-enabled SC model, as shown in Fig. \ref{fig:semcom}, which utilizes LLMs as the semantic decoder, leveraging extensive prior knowledge and adaptive reasoning capabilities to enhance interpretation for time-varying channels. Specifically, a  semantic encoder, typically implemented using convolutional neural networks (CNNs), is employed to extract high-level semantic representations from source data originating from the SAGSIN\footnote{The CNN encoder reduces each 512$\times$512 image to a 256-dimensional feature vector, roughly 0.2\% of the raw size, using a ResNet-like backbone.}. These semantic features are subsequently passed through a channel encoder to perform digital-to-analog conversion, facilitating wireless transmission\footnote{The physical-layer signal processing chain mirrors the spirit of distributed source–channel coding: LDPC or BCH codes protect the semantic bits and 16-QAM maps them to complex-valued symbols, as in B.\ Girod, ``Distributed Source Coding: A Statistical Theory Perspective,'' \textit{Proc.\ IEEE}, 2005. This is why the channel-encoder of Fig. \ref{fig:semcom} carries out A/D-conversion and constellation-mapping.}.
	Following propagation through a noisy communication channel, the received signal is processed by a channel decoder to recover the encoded semantic representations, with the restoration quality often visualized via constellation diagrams\footnote{A conventional demapper and LDPC (or BCH) decoder undoes the modulation and channel coding, producing a potentially noisy 256-dimensional semantic vector.}.
	The core enhancement of the system lies in the integration of a powerful semantic decoder powered by large language models (LLMs) such as LLaMA, GPT, and DeepSeek with an underwater communication channel. These LLMs possess advanced reasoning and comprehension capabilities, allowing them to infer meaning from partial or degraded semantic inputs\footnote{For image reconstruction the LLM inserts missing pixels by 'hallucination'; for classification (e.g., detecting fish in subsea imagery) it outputs categorical labels such as ``present'' or ``absent.''}.
	By leveraging the rich contextual understanding and strong generalization capability of LLMs, the semantic decoder accurately performs task-specific functions such as reconstruction, localization, detection, and classification. 
	
	This LLM-driven SC model significantly enhances the performance, especially in complex heterogeneous SAGSIN environments.
	
	\subsection{Experimental Settings}
	We consider a typical coastal communication scenario between autonomous underwater vehicles (AUVs) and surface platforms. Specifically, the link range is set to 2 km with a water depth of $H=50$ m, representing medium-depth coastal waters. A center frequency of 12 kHz and a bandwidth of 3 kHz are chosen in alignment with commercial underwater modems. 
	The number of resolvable multipath components is determined by geometry, yielding $N = \lceil 1 + 2 H f_c/c \rceil = 6$, where $f_c$ is the center frequency, and $c = 1500$ m/s denotes the speed of sound in water. 
	The line-of-sight delay is calculated as $\tau_0 = d/c = 1.33$ s. The LoS tap is modeled using a Rician distribution with $K=6$ dB, while scattered multipath components follow a Rayleigh distribution. Doppler effects are modelled by uniformly distributed Doppler shifts in the range $[-f_{D,\max}, f_{D,\max}]$, with $f_{D,\max}=v_\text{rel}f_c/c=12$ Hz, assuming a relative velocity of 1.5 m/s. The resultant phase evolution is modelled as a linear function of time, and slow variations in path delays are introduced through a Gaussian perturbation with standard deviation of $\sigma_\tau = 50~\mu s$ to emulate surface motion dynamics.
	For the SC model, the semantic encoder is constructed based on CNNs, and the semantic decoder is based on LLaMA3 \cite{grattafiori2024llama}.
	
	We adopt the Seaclear Marine Debris Dataset \cite{djuravs2024dataset} as the training data, which contains 8,610 images of marine litter, encompassing not only litter but also animals, plants, and robot parts observed. The SC model aims for ensuring the consistency between the raw and received images by only transmitting the semantic features. Hence, the Structural Similarity Index (SSIM) is used as the evaluation metric. 
	
	\subsection{Performance Results}
\begin{figure}[htbp]
		\centering
		\includegraphics[width=6cm]{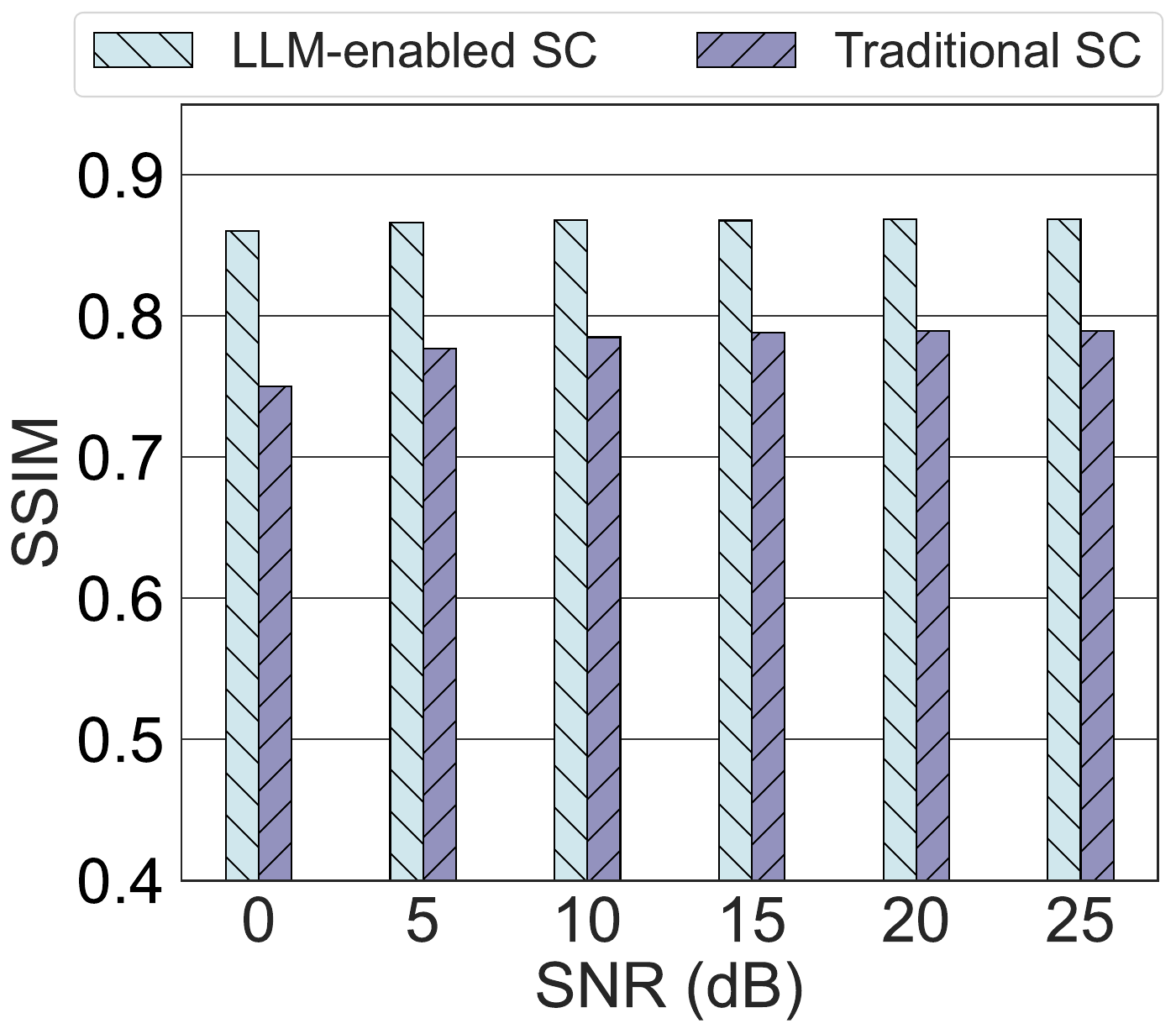}
		\caption{The SSIM results between the LLM-enabled SC and a conventional DeepSC baseline (CNN encoder + GRU decoder).}
		\label{fig:exp}
	\end{figure}

	As illustrated in Fig. \ref{fig:exp}, the proposed LLM-assisted SC system consistently outperforms the traditional SC baseline across a wide range of SNRs. In particular, the SSIM achieved by the LLM-based method exhibits a clear upward trend upon increasing the SNR and maintains a significant margin over the traditional approach at all tested points. The traditional baseline follows the widely studied DeepSC architecture \cite{Xie2021}, which employs a CNN-based semantic encoder and a lightweight gated recurrent unit (GRU) based decoder without any large-parameter language model.
	
	This performance gain highlights the effectiveness of integrating LLMs into the SC framework. By leveraging their powerful semantic representation and context modelling capabilities, LLMs accurately capture and reconstruct the essential meaning of transmitted information, even under noisy channel conditions.
	
	\section{Open Challenges \& Future Directions}
	
	\subsection{LLM-Assisted Channel Prediction}
	
	Before large-parameter transformers may become routine CSI engines for SAGSIN, three hurdles must be cleared.
	
	\subsubsection*{On-device efficiency}
	Satellites and buoy gateways operate on single-digit-Watt power budgets; even 7-Billion parameter based LLama-2/3 models - which are considered compact - have excessive memory and multiply–accumulate counts. Sub-4-bit quantisation, low-rank adaptation, and task-specific distillation tuned to physical-layer inputs remain open research areas.
	
	\subsubsection*{Domain shift and data scarcity}
	Public LEO, HAP, and underwater channel traces are scarce. Models pretrained on synthetic corpora must adapt online without catastrophic forgetting. Federated continual learning that passes gradients—rather than raw IQ samples—across SAGSIN tiers is largely unexplored.
	
	\subsubsection*{Reliability and interpretability}
	A mis-predicted channel may mis-align a beam and trigger sudden outage, hence operators must know how and why an LLM created its forecast. As a remedy, interpretable surrogates, such as attention-weight heat-maps over the delay–Doppler grid, Shapley-value attributions on tokenised channel patches, or physics-guided consistency checks may be used to flag violations of power conservation. Explicitly, they can expose which specific paths, ports, or time windows drive each decision. Coupled with carefully calibrated confidence scores and a skip-prediction fallback, these tools bound worst-case error, while offering valuable insights for debugging and online adaptation.
	
	\subsection{LLM-Enabled Semantic Communication}
	
	The semantic pipeline presented in Section~\ref{section:SC} handles single-modality imagery; practical services require more flexibility.
	
	\subsubsection*{Multimodal fusion}
	Future SAGSIN packets will blend text commands, LiDAR point clouds, sonar spectrograms and high-resolution imagery.  Du \textit{et al.}~\cite{Du2025} demonstrate that a large multimodal model can learn cross-modal saliency cues and adjust transmission priorities in vehicular networks. However, an equivalent framework for satellite–air–sea channels is still missing.  A foundation model must therefore fuse heterogeneous tensors, while allocating source and error protection bits unequally across modalities, all within the tight SNR budgets of non-terrestrial links.
	
	\subsubsection*{Task-configurable fidelity}
	Remote inspection prefers structural similarity, while target recognition highly values object integrity. Encoders must exhibit fidelity/bit-rate flexibility without full retraining.
	
	\subsubsection*{Ultra-low-latency decoding}
	UAV relay swarms impose end-to-end delays below 20~ms. Lightweight attention or hierarchical token streaming is needed to meet control-loop deadlines.
	\subsubsection*{Evaluation metrics}
	Traditional fidelity metrics such as peak SNR (PSNR) or SSIM only measure pixel-level similarity to the original image; they say nothing about whether the decoded payload still enables the downstream task so a packet could have high SSIM yet be useless for the mission. Unified metrics that correlate with mission outcomes across radio, optical, and acoustic links remain to be standardised.
	
	\subsection{Cross-Layer Resource Allocation and Control}
	
	Channel predictors and semantic encoders ultimately feed beam selection, power control, spectrum scheduling and multi-hop routing, yet today these functions are tuned in isolation and on different time scales.  Yang \textit{et al.}~\cite{Yang2025} show that a foundation model pretrained on diverse wireless traces can be prompted to handle multiple link-layer tasks, suggesting that a single LLM agent could forecast CSI, pick beams, set compression ratios and even call network simulators in a single inference pass.  Designing reward structures, latency budgets and safety guards for such composite policies therefore remains an open frontier.
	
	\subsection{System-Level Considerations}
	
	Beyond algorithms, several system-level hurdles stand between prototypes and deployment.
	
	\subsubsection*{Model governance and security}
	Adversarial delay–Doppler tokens or semantic streams could bias network behaviour. Robustness certificates and signed provenance logs are needed for on-orbit model updates.
	
	\subsubsection*{Benchmarking and standardisation}
	No shared data set spans SAGSIN’s radio, optical and acoustic channels alongside application-layer KPIs.  The ETSI Technology Radar report on AI for non-terrestrial networks highlights the urgent need for open test suites and common data repositories.  Early liaison among ETSI ISG-NTN, CCSDS and emerging IEEE working groups is essential to ensure interoperability.

	\subsubsection*{Regulatory alignment}
	Dynamic spectrum reuse by LEO mega-constellations is already contentious; adaptive, learning-based layers will require transparent audit trails to satisfy spectrum and maritime regulators.
	
	\medskip
	Solving these interlocking challenges—from efficient LLM compression and multimodal semantics to cross-layer control and governance—will decide how quickly SAGSIN evolves from concept to global reality. Looking ahead, a plausible deployment scenario by 2030 might feature onboard inference engines hosted directly by LEO satellites and offshore buoys. These systems could run 4-bit quantized LLMs, preloaded with physics-informed token embeddings and be continually refined through secure over-the-air updates using LoRA-style patching. Ground stations could serve as federation hubs, aggregating gradients from across the SAGSIN stack and redistributing distilled updates without transmitting raw data. By incorporating trustworthy execution environments and blockchain-audited update logs, such architectures could enable resilient, self-improving communication layers that operate autonomously across space, air, ground and sea scenarios.

	\section{Conclusion}

	LLMs were shown to tackle two persistent hurdles in SAGSINs: CSI-aging and severe bandwidth imbalance. A fluid-antenna based LLM predictor forecasts delay–Doppler channels several coherence intervals ahead, while an LLM-based semantic encoder turns raw sensor data into task-aware tokens, cutting SNR needs by $> 10 $dBs. Together they form a medium-agnostic, data-driven layer that spans radio, optical and acoustic links.
	
	The next key steps are lighter models for on-device use, multimodal fidelity control, unified cross-layer decision engines and open benchmarks for trustworthy operation. Progress along these lines will decide how promptly LLM-assisted adaptation moves from lab demos to field deployment—and whether SAGSIN can realise its promise of truly ubiquitous connectivity.
	

\bibliographystyle{IEEEtran}
\bibliography{reference}
	
\end{document}